# Digital Signal Processing for Optical Communications and Networks I:

# Linear Compensation

Tianhua Xu (Dr.)

Department of Electronic and Electrical Engineering, University College London, London, United Kingdom

**Abstract:** The achievable information rates of optical communication networks have been widely increased over the past four decades with the introduction and development of optical amplifiers, coherent detection, advanced modulation formats, and digital signal processing techniques. These developments promoted the revolution of optical communication systems and the growth of Internet, towards the direction of high-capacity and long-distance transmissions. The performance of long-haul high-capacity optical fiber communication systems is significantly degraded by transmission impairments, such as chromatic dispersion, polarization mode dispersion, laser phase noise and Kerr fiber nonlinearities. With the entire capture of the amplitude and phase of the signals using coherent optical detection, the powerful compensation and effective mitigation of the transmission impairments can be implemented using the digital signal processing in electrical domain. This becomes one of the most promising techniques for next-generation optical communication networks to achieve a performance close to the Shannon capacity limit.

Chromatic dispersion can be compensated using the digital filters in both time domain and frequency domain. Polarization mode dispersion can be equalized adaptively using the least-mean-square method and constant modulus algorithm. Phase noise from laser sources can be estimated and compensated using the feed-forward and feed-back carrier phase estimation approaches. Kerr fiber nonlinearities, including self-phase modulation, cross-phase modulation and four-wave mixing, can be mitigated using single-channel digital back-propagation and multi-channel digital back-propagation, where both intra-channel and inter-channel nonlinearities can be compensated. Digital signal processing combined with coherent detection shows a very promising solution for long-haul high capacity optical fiber communication systems, which offers a great flexibility in the design, deployment, and operation of optical communication networks. This chapter will focus on the introduction and investigation of digital signal processing employed for channel impairments compensation based on the coherent detection of optical signals, to provide a roadmap for the design and implementation of real-time optical fiber communication systems.

**Keywords:** Optical communications, Optical networks, Digital signal processing, Coherent detection, Chromatic dispersion, Polarization mode dispersion, Laser phase noise, Fiber nonlinearities

# 1. Introduction

The performance of high-capacity optical communication systems can be significantly degraded by fiber attenuation, chromatic dispersion (CD), polarization mode dispersion (PMD), laser phase noise (PN) and Kerr nonlinearities [1-10]. Using coherent detection, the powerful compensation of transmission impairments can be implemented in electrical domain. With the full information of the received signals, the chromatic dispersion, the polarization mode dispersion, the carrier phase noise and the fiber Kerr nonlinearities can be equalized and mitigated using digital signal processing (DSP) [11-22].

Due to the high sensitivity of the receiver, coherent optical transmission was investigated extensively in the eighties of last century [23,24]. However, the development of coherent communication has been delayed for nearly 20 years after that period [25,26]. Coherent optical detection re-attracted the research interests until 2005, since the advanced modulation formats i.e. $m$-level phase shift keying ($m$-PSK) and $m$-level quadrature amplitude modulation ($m$-QAM) can be applied [27-30]. In addition, coherent optical detection allows the electrical mitigation of system impairments. With the two main merits, the reborn coherent detections brought us the enormous potential for higher transmission speed and spectral efficiency in current optical fiber communication systems [31,32].

With an additional local oscillator (LO) source, the sensitivity of coherent receiver reached the limitation of the shot-noise. Furthermore, compared to the traditional intensity modulation direct detection system, the multi-level modulation formats can be applied using the phase modulations, which can include more information bits in one transmitted symbols than before.

Meanwhile, since the coherent demodulation is linear and all information of the received signals can be detected, signal processing approaches i.e. tight spectral filtering, CD equalization, PMD compensation, laser PN estimation, and fiber nonlinearity compensation can be implemented in electrical domain [33-40].

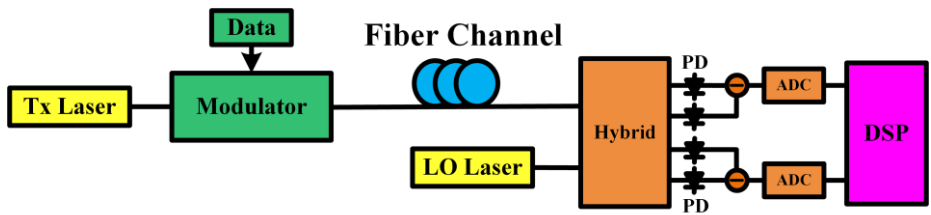

Figure 1. Schematic of coherent optical communication system with digital signal processing.

The typical block diagram of the coherent optical transmission system is shown in Figure 1. The transmitted optical signal is combined coherently with the continuous wave from the narrow-linewidth LO laser, so that the detected optical intensity in the photodiode (PD) ends can be increased and the phase information of the optical signal can be obtained. The use of LO laser is to increase the receiver sensitivity of the detection of optical signals, and the performance of coherent transmission can even behave close to the Shannon limit [3,12].

The development of the coherent transmission systems has stopped for more than 10 years due to the invention of Erbium-doped fiber amplifiers (EDFAs) [1,2]. The coherent transmission techniques attracted the interests of investigation again around 2005, when a new stage of the coherent lightwave systems comes out by combining the digital signal processing techniques [41-46]. This type of coherent lightwave system is called as digital coherent communication system. In the digital coherent transmission systems, the electrical signals output from the photodiodes are sampled and transformed into the discrete signals using high-speed analogue-to-digital convertors (ADCs), which can be further processed by the DSP algorithms.

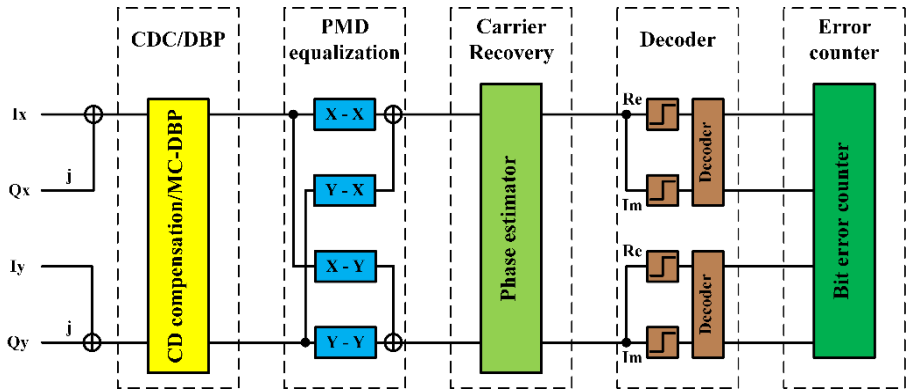

Figure 2: Block diagram of DSP in digital coherent receiver.

The phase locking and the polarization adjustment were the main obstacles in the traditional coherent lightwave systems, while they can be solved by the carrier phase estimation and the polarization equalization respectively in the digital coherent optical transmission systems [47-55]. Besides, the chromatic dispersion and the nonlinear effects can also be mitigated by using the digital signal processing techniques [56-62]. The typical structure of the DSP compensating modules in the digital coherent receiver is shown in Figure 2.

## 2. Digital signal processing for compensating transmission impairments

In this section, the chromatic dispersion compensation, polarization mode dispersion equalization and carrier phase noise compensation are analyzed and discussed using corresponding DSP algorithms.

## 2.1 Chromatic dispersion compensation

Digital filters involving the time-domain least-mean-square (TD-LMS) adaptive filter, the static time-domain finite impulse response (STD-FIR) filter, and the frequency-domain equalizers (FDEs) are investigated for CD compensation. The characters of these filters are analyzed based on a 28-Gbaud dual-polarization quadrature phase shift keying (DP-QPSK) coherent transmission system using post-compensation of dispersion. It is noted that the STD-FIR filter and the FDEs can also be used for the dispersion pre-distorted coherent communication systems.

### 2.1.1 Time domain least-mean-square equalizer

The TD-LMS filter employs an iterative algorithm that incorporates successive corrections to weights vector in the negative direction of the gradient vector which eventually leads to a minimum mean square error [34,38,63-65]. The transfer function of the TD-LMS digital filter can be described as follows:

$$y_{out}(n) = \vec{W}_{LMS}^{H}(n)\vec{x}_{in}(n) \tag{1}$$

$$\vec{W}_{LMS}(n+1) = \vec{W}_{LMS}(n) + \mu_{LMS}\vec{x}_{in}(n)e_{LMS}^{*}(n) \tag{2}$$

$$e_{LMS}(n) = d_{LMS}(n) - y_{out}(n) \tag{3}$$

where $\vec{x}_{in}(n)$ is the vector of received signals, $y_{out}(n)$ is the equalized output signal, $n$ is the index of signal, $\vec{W}_{LMS}(n)$ is the vector of tap weights, $H$ is the Hermitian transform operator, $d_{LMS}(n)$ is the desired symbol, $e_{LMS}(n)$ is the error between the desired symbol and the output signal, * is the conjugation operator, and $\mu_{LMS}$ is the step size. To ensure the convergence of tap weights $\vec{W}_{LMS}(n)$, the step size $\mu_{LMS}$ has to meet the condition of $\mu_{LMS} < 1/U_{max}$, where $U_{max}$ is the largest eigenvalue of the correlation matrix $R = \vec{x}_{in}(n)\vec{x}_{in}^{H}(n)$ [63]. The TD-LMS dispersion compensation filter can be applied in the "decision-directed" or the "sequence-training" mode [63].

The tap weights in TD-LMS adaptive equalizer for 20 km fiber CD compensation is shown in Figure 3. The convergence for 9 tap weights in the TD-LMS filter with step size equal to 0.1 is shown in Figure 3(a), and it is found that the tap weights reach their convergence after ~5000 iterations. The distribution of the magnitudes of the converged tap weights is plotted in Figure 3(b), and it is found that the central tap weights take more dominant roles than the high-order tap weights [34,66].

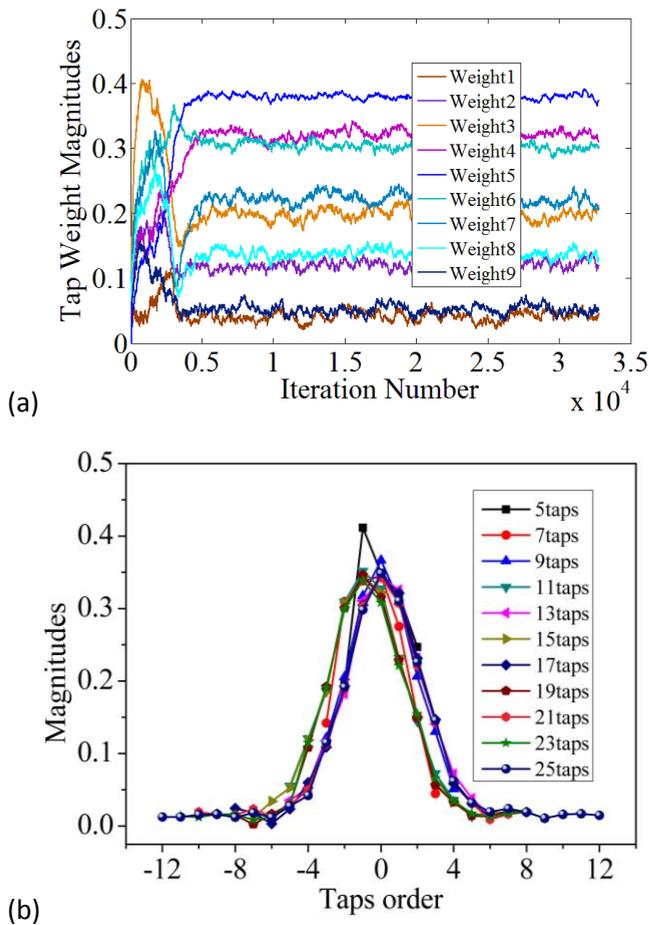

(a)

(b)

Figure 3: Taps weights of TD-LMS filter. (a) Tap weights magnitudes convergence. (b) Converged tap weights magnitudes distribution.

### 2.1.2 Static time-domain finite impulse response filter

Compared with the iteratively updated TD-LMS filter, the tap weights in STD-FIR filter have a relatively simple specification [34,67-69], the tap weight in STD-FIR filter is given by the following equations:

$$a_k = \sqrt{\frac{jcT^2}{D\lambda^2 L}} \exp\left(-j\frac{\pi c T^2}{D\lambda^2 L}k^2\right) \quad -\left\lfloor\frac{N}{2}\right\rfloor \leq k \leq \left\lfloor\frac{N}{2}\right\rfloor \quad (4)$$

$$N^A = 2\times\left\lfloor\frac{|D|\lambda^2 L}{2cT^2}\right\rfloor + 1 \quad (5)$$

where $D$ is the CD coefficient, $\lambda$ is the carrier central wavelength, $L$ is the length of fiber, $T$ is the sampling period, $N^A$ is the maximum number of taps, and $\lfloor x \rfloor$ means the nearest integer smaller than $x$.

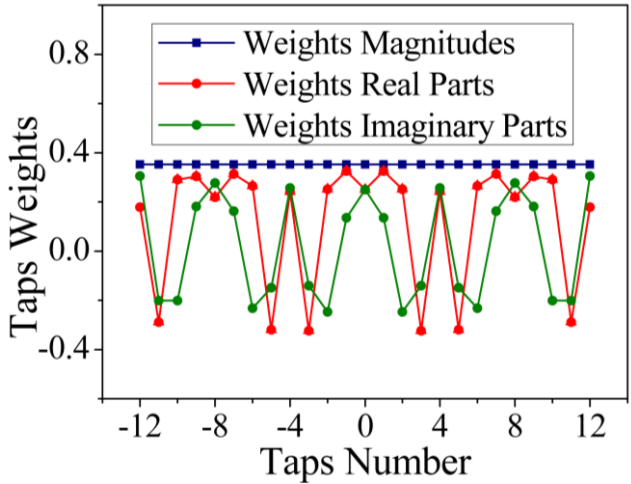

Figure 4: Tap weights of STD- FIR chromatic dispersion compensation filter.

For 20 km fiber with CD coefficient of $D=16\,ps/(nm\cdot km)$, the distribution of the tap weights in the STD-FIR filter is shown in Figure 4.

### 2.1.3 Frequency domain equalizers

Since the complexity is very low for compensating large CD [34,70], the most promising and popular chromatic dispersion compensation filters in coherent transmission systems are the frequency domain equalizers. The transfer function of the frequency domain equalizers is given by the following expression:

$$G_c(L,\omega) = \exp\left(\frac{-jD\lambda^2\omega^2 L}{4\pi c}\right) \quad (6)$$

where $D$ is the chromatic dispersion coefficient, $\lambda$ is the carrier central wavelength, $\omega$ is the angular frequency, $L$ is the length of fiber, and $c$ is the light speed in vacuum.

The frequency domain equalizers are generally implemented using the overlap-save (OLS) and the overlap-add (OLA) approaches based on the fast Fourier transform and the inverse fast Fourier transform (iFFT) convolution algorithms [71-73], as described in Figure 5.

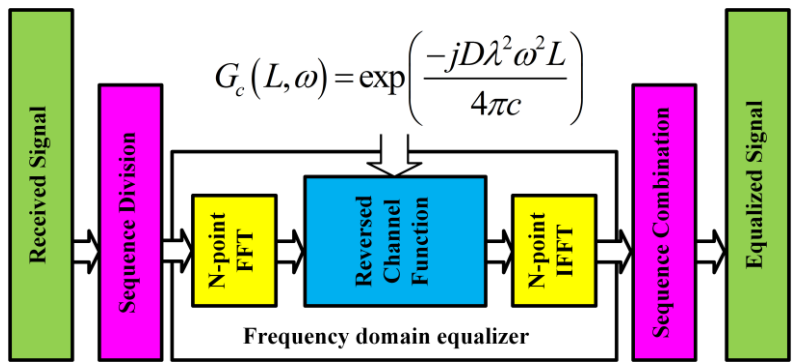

Figure 5: Schematic of frequency domain equalizer for chromatic dispersion compensation.

### 2.2 Polarization mode dispersion equalization

Due to the random character of the polarization mode dispersion and the polarization rotation, the compensation of the PMD and the polarization rotation is generally realized by the adaptive algorithms such as the least-mean-square (LMS) and the constant modulus algorithm (CMA) filters.

### 2.2.1 LMS adaptive PMD equalization

In the electrical domain, the impact of the PMD and the polarization fluctuation can be adaptively equalized using the decision-directed LMS (DD-LMS) filter [36,63], of which the transfer function is given by:

$$\begin{bmatrix} x_{out}(n) \\ y_{out}(n) \end{bmatrix} = \begin{bmatrix} \vec{w}_{xx}^{H}(n) & \vec{w}_{xy}^{H}(n) \\ \vec{w}_{yx}^{H}(n) & \vec{w}_{yy}^{H}(n) \end{bmatrix} \cdot \begin{bmatrix} \vec{x}_{in}(n) \\ \vec{y}_{in}(n) \end{bmatrix} \quad (7)$$

$$\begin{cases} \vec{w}_{xx}(n+1) = \vec{w}_{xx}(n) + \mu_p \cdot \varepsilon_x(n) \cdot \vec{x}_{in}^{*}(n) \\ \vec{w}_{yx}(n+1) = \vec{w}_{yx}(n) + \mu_p \cdot \varepsilon_y(n) \cdot \vec{x}_{in}^{*}(n) \\ \vec{w}_{xy}(n+1) = \vec{w}_{xy}(n) + \mu_p \cdot \varepsilon_x(n) \cdot \vec{y}_{in}^{*}(n) \\ \vec{w}_{yy}(n+1) = \vec{w}_{yy}(n) + \mu_p \cdot \varepsilon_y(n) \cdot \vec{y}_{in}^{*}(n) \end{cases} \quad (8)$$

$$\begin{cases} \varepsilon_x(n) = d_x(n) - x_{out}(n) \\ \varepsilon_y(n) = d_y(n) - y_{out}(n) \end{cases} \quad (9)$$

where $\vec{x}_{in}(n)$ and $\vec{y}_{in}(n)$ are the vectors of the input signals, $x_{out}(n)$ and $y_{out}(n)$ are the equalized output signals respectively, $\vec{w}_{xx}(n)$, $\vec{w}_{xy}(n)$, $\vec{w}_{yx}(n)$ and $\vec{w}_{yy}(n)$ are the complex tap weights vectors, $d_x(n)$ and $d_y(n)$ are the desired symbols, $\varepsilon_x(n)$ and $\varepsilon_y(n)$ are the estimation errors between the desired symbols and the output signals in the two polarizations, respectively, and $\mu_p$ is the step size in the DD-LMS algorithm.

### 2.2.2 CMA adaptive PMD equalization

The influence of the PMD and the polarization fluctuation can also be compensated employing the CMA adaptive filter [74,75], of which the transfer function can be described as:

$$\begin{bmatrix} x_{out}(n) \\ y_{out}(n) \end{bmatrix} = \begin{bmatrix} \vec{v}_{xx}^H(n) & \vec{v}_{xy}^H(n) \\ \vec{v}_{yx}^H(n) & \vec{v}_{yy}^H(n) \end{bmatrix} \cdot \begin{bmatrix} \vec{x}_{in}(n) \\ \vec{y}_{in}(n) \end{bmatrix} \qquad (10)$$

$$\begin{cases} \vec{v}_{xx}(n+1) = \vec{v}_{xx}(n) + \mu_q \cdot \eta_x(n) \cdot \vec{x}_{in}^*(n) \\ \vec{v}_{yx}(n+1) = \vec{v}_{yx}(n) + \mu_q \cdot \eta_y(n) \cdot \vec{x}_{in}^*(n) \\ \vec{v}_{xy}(n+1) = \vec{v}_{xy}(n) + \mu_q \cdot \eta_x(n) \cdot \vec{y}_{in}^*(n) \\ \vec{v}_{yy}(n+1) = \vec{v}_{yy}(n) + \mu_q \cdot \eta_y(n) \cdot \vec{y}_{in}^*(n) \end{cases} \qquad (11)$$

$$\begin{cases} \eta_x(n) = 1 - |x_{out}(n)|^2 \\ \eta_y(n) = 1 - |y_{out}(n)|^2 \end{cases} \qquad (12)$$

where $\vec{x}_{in}(n)$ and $\vec{y}_{in}(n)$ are the vectors of the input signals, $x_{out}(n)$ and $y_{out}(n)$ are the equalized output signals respectively, $\vec{v}_{xx}(n)$, $\vec{v}_{xy}(n)$, $\vec{v}_{yx}(n)$ and $\vec{v}_{yy}(n)$ are the complex tap weights vectors, $\eta_x(n)$ and $\eta_y(n)$ are the estimation errors between the desired amplitude and the output signals in the two polarizations, respectively, and $\mu_q$ is the step size in the CMA algorithm.

It can be found that the CMA algorithm is based on the principle of minimizing the modulus variation of the output signal to update its weight vector.

### 2.3 Carrier phase estimation

In this section, the analyses on different carrier phase estimation algorithms, involving the one-tap normalized LMS, the differential phase estimation, the block-wise average (BWA) and the Viterbi-Viterbi (VV) methods in the coherent optical transmission systems will be presented.

### 2.3.1 The normalized LMS carrier phase estimation

The one-tap normalized LMS filter can be employed effectively for carrier phase estimation [76-78], of which the tap weight is expressed as

$$w_{NLMS}(n+1) = w_{NLMS}(n) + \frac{\mu_{NLMS}}{|x_{in}(n)|^2} x_{in}^*(n) e_{NLMS}(n) \tag{13}$$

$$e_{NLMS}(n) = d_{PE}(n) - w_{NLMS}(n) \cdot x_{in}(n) \tag{14}$$

where $w_{NLMS}(n)$ is the tap weight, $x_{in}(n)$ is the input signal, $n$ is the symbol index, $d_{PE}(n)$ is the desired symbol, $e_{NLMS}(n)$ is the carrier phase estimation error between the desired symbol and the output signal, and $\mu_{NLMS}$ is the step size in the one-tap normalized LMS filter.

It has been demonstrated that the one-tap normalized LMS carrier phase estimation behaves similar to the differential phase estimation [28,53,55,76], of which the BER floor in the $m$-PSK coherent optical transmission systems can be approximately described by the following analytical expression,

$$BER_{floor}^{NLMS} \approx \frac{1}{\log_2 m} erfc\left(\frac{\pi}{m\sqrt{2}\sigma}\right) \tag{15}$$

where $\sigma$ is the square root of the phase noise variance. The schematic of the one-tap normalized LMS carrier phase estimation is illustrated in Figure 6.

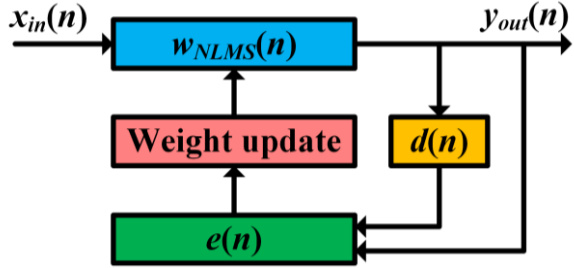

Figure 6: Schematic of one-tap normalized LMS carrier phase estimation.

### 2.3.2 Differential carrier phase estimation

The differential signal demodulation can also be applied for carrier phase estimation in coherent transmission system [28,53,55], where the differentially encoded data can be recovered using the "delay and multiply" algorithm. Using differential carrier phase estimation, the encoded information can be recovered according to the phase difference between the two consecutive symbols, i.e. the decision variable $\Psi = x_n x_{n+1}^* \exp\{i\pi/m\}$, where $x_n$ and $x_{n+1}$ are the consecutive n-th and (n+1)-th received symbols. The BER floor of the differential carrier phase estimation can be evaluated using the principle of conditional probability. For the *m*-PSK coherent systems, the BER floor in differential phase estimation is expressed as the following equation [28,53],

$$BER_{floor}^{Differential} = \frac{1}{\log_2 m} erfc\left(\frac{\pi}{m\sqrt{2}\sigma}\right) \qquad (16)$$

where σ is the square root of the phase noise variance. The schematic of the differential carrier phase estimation is described in Figure 7.

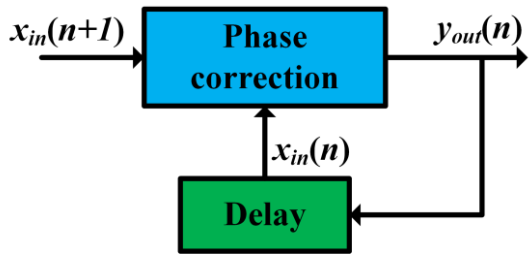

Figure 7: Schematic of differential carrier phase estimation.

### 2.3.3 The block-wise average carrier phase estimation

The block-wise average approach calculates the *m*-th power of the received symbols in each processing unit to remove the information of phase modulation, and the computed phase are summed and averaged over the entire process block, where the length of the process block is called block size. Then the averaged phase is divided by *m*, and the result leads to the phase estimate for the entire data block [79-81]. For the *m*-PSK coherent communication system, the estimated carrier phase in each process block using the block-wise average approach is given by the following expression:

$$\hat{\Phi}_{BWA}(n) = \frac{1}{m}\arg\left\{\sum_{k=1+(M-1)\cdot N_b}^{M\cdot N_b} x^m(k)\right\} \quad (17)$$

$$M = \left\lceil \frac{n}{N_b} \right\rceil \quad (18)$$

where $N_b$ is the block size in the BWA approach, and $\lceil x \rceil$ means the nearest integer larger than $x$.

The performance of the block-wise average carrier phase estimation method in the $m$-PSK coherent optical communication system can be derived based on the Taylor expansion of the estimated carrier phase error, and the BER floor in the block-wise average carrier phase estimation can be described using the following expression [52,53,55,79]:

$$BER_{floor}^{BWA} \approx \frac{1}{N_b \cdot \log_2 m} \cdot \sum_{k=1}^{N_b} erfc\left(\frac{\pi}{m\sqrt{2}\sigma_{BWA,k}}\right) \quad (19)$$

$$\sigma_{BWA,k}^2 = \frac{\sigma^2}{6N_b^2} \cdot \left[2(k-1)^3 + 3(k-1)^2 + 2(N_b-k)^3 + 3(N_b-k)^2 + N_b - 1\right] \quad (20)$$

where $\sigma^2$ represents the total phase noise variance in the coherent transmission system. The schematic of the block-wise average carrier phase estimation is shown in Figure 8.

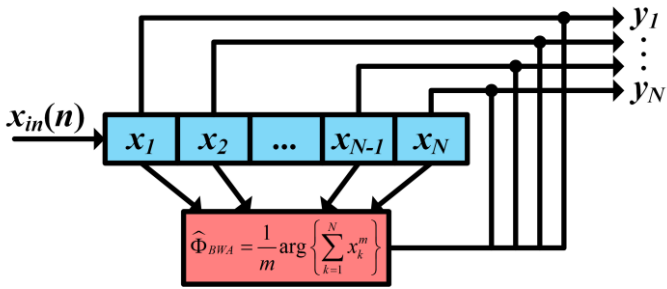

Figure 8: Schematic of block-wise average carrier phase estimation.

### 2.3.4 The Viterbi-Viterbi carrier phase estimation

The Viterbi-Viterbi carrier phase estimation approach also operates the symbols in each process block into the *m*-th power to remove the information of the phase modulation. The computed phase are also summed and averaged over the entire process block, where the length of the process block is also called block size. Then the averaged phase is divided by *m* as the estimated carrier phase. However, compared to the BWA approach, the estimated phase in the Viterbi-Viterbi carrier phase estimation approach is only applied in the phase recovery of the central symbol in each process block [55,81-83]. The estimated carrier phase in the Viterbi-Viterbi approach in *m*-PSK optical communication systems is given by the following expression:

$$\hat{\Phi}_{VV}(n) = \frac{1}{m}\arg\left\{\sum_{k=-(N_v-1)/2}^{(N_v-1)/2} x^m(n+k)\right\}, \quad N_v=1,3,5,7... \quad (21)$$

where $N_v$ is the block size in the Viterbi-Viterbi carrier phase estimation approach.

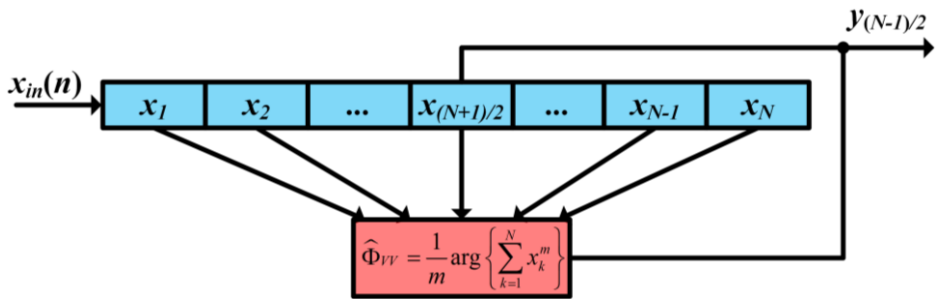

Figure 9: Schematic of Viterbi-Viterbi carrier phase estimation.

The performance of the Viterbi-Viterbi carrier phase estimation in the *m*-PSK coherent optical communication system can also be derived employing the Taylor expansion of the estimated carrier phase. The BER floor in the Viterbi-Viterbi carrier phase estimation for the *m*-PSK transmission system can be expressed as follows [52,53,55]:

$$BER_{floor}^{VV} \approx \frac{1}{\log_2 m} erfc\left(\frac{\pi}{m\sqrt{2}\sigma_{VV}}\right) \quad (22)$$

$$\sigma_{VV}^2 = \sigma^2 \cdot \frac{N_v^2 - 1}{12 N_v} \tag{23}$$

where $\sigma^2$ represents the total phase noise variance in the coherent transmission system. The schematic of the Viterbi-Viterbi carrier phase estimation is illustrated in Figure 9.

According to Eq. (20) and Eq. (23), it can be found that the phase estimate error in the Viterbi-Viterbi carrier phase estimation corresponds to the phase estimate error of the central symbol (the smallest error) in the block-wise average carrier phase estimation. Therefore, the Viterbi-Viterbi approach will generally performs better than the block-wise average approach, in term of the phase estimate error. However, it requires more computational complexity to update the process unit for the phase estimation of each symbol.

It is noted that the one-tap normalized LMS algorithm can also be employed for the $m$-QAM coherent transmission systems, while the block-wise average and the Viterbi-Viterbi methods cannot be easily used for the classical $m$-QAM coherent systems except the circular constellation $m$-QAM systems.

## 3. Conclusions

In this chapter, the digital signal processing techniques for compensating transmission impairments in optical communication systems, including chromatic dispersion, polarization mode dispersion and laser phase noise have been described and analyzed in detail. Chromatic dispersion can be compensated using the digital filters in both time domain and frequency domain. Polarization mode dispersion can be equalized adaptively using the least-mean-square method and the constant modulus algorithm. Phase noise from the laser sources can be estimated and compensated using the feed-forward and feed-back carrier phase recovery approaches.

Digital signal processing combined with coherent detection shows a very promising solution for long-haul high-capacity optical communication systems, which offers a great flexibility in the design, deployment, and operation of optical communication networks. Fiber nonlinearities, including self-phase modulation, cross-phase modulation and four-wave mixing, can be mitigated using single-channel and multi-channel digital back-propagation in the electrical domain, which will be discussed in future work.

## 4. Acknowledgements

This work is supported in part by UK Engineering and Physical Sciences Research Council (project UNLOC EP/J017582/1), in part by European Commission Research Council FP7-PEOPLE-2012-IAPP (project GRIFFON, No. 324391), in part by European Commission Research Council FP7-PEOPLE-2013-ITN (project ICONE, No. 608099), and in part by Swedish Research Council Vetenskapsradet (No. 0379801).

## 5. References

[1] G. P. Agrawal, Fiber-optic communication systems (4th Edition), John Wiley & Sons, Inc., 2010. ISBN: 978-0-470-50511-3
[2] I. Kaminow, T. Li, A.E. Willner, Optical Fiber Telecommunications VB: System and Networks (5th Edition), Academic Press, Oxford, 2010. ISBN: 978-0-12-374172-1
[3] D. Semrau, et al., Achievable information rates estimates in optically amplified transmission systems using nonlinearity compensation and probabilistic shaping, Optics Letters, Vol.41(21), 121-124, 2017. DOI: 10.1364/OL.42.000121
[4] Y. Li, et al., Dynamic dispersion compensation in a 40 Gb/s single-channeled optical fiber communication system, ACTA OPTICA SINICA, Vol.27, 1161-1165, 2007. ISSN: 0253-2239.2007.07.004
[5] T. Xu, et al., Overcoming fibre nonlinearities to enhance the achievable transmission rates in optical communication systems, Asia Communications and Photonics Conference, Workshop 3, 2015. (Invited Talk)
[6] G. Liga, et al., Ultra-wideband nonlinearity compensation performance in the presence of PMD, European Conference on Optical Communication, 794-796, 2016. ISBN: 978-3-8007-4274-5
[7] T. Xu, DSP based chromatic dispersion equalization and carrier phase estimation in high speed coherent optical transmission systems, KTH Ph.D. Thesis, 2012. ISBN 978-91-7501-346-6
[8] S. Sergeyev, et al., All-optical polarisation control in fibre Raman amplifiers, International Scientific and Technical Conference on Quantum Electronics, 81, 2013. ISBN: 978-985-553-157-0
[9] G. Jacobsen, et al., Phase noise influence in coherent optical DnPSK systems with DSP based dispersion compensation, Journal of Optical Communications, Vol.35(1), 57-61, 2014. DOI: 10.1515/joc-2013-0065

[10] T. Xu, et al., Field trial over 820 km installed SSMF and its potential Terabit/s superchannel application with up to 57.5-Gbaud DP-QPSK transmission, Optics Communications, Vol.353, 133-138, 2015. DOI: 10.1016/j.optcom.2015.05.029
[11] E. Ip, A. P. T. Lau, D. J. F. Barros, J. M. Kahn, Coherent detection in optical fiber systems, Optics Express, Vol. 16(2), 753-791, 2008. DOI: 10.1364/OE.16.000753
[12] G. Li, Recent advances in coherent optical communication, Advances in Optics and Photonics, Vol. 1(2), 279-307, 2009. DOI: 10.1364/AOP.1.000279
[13] G. Jacobsen, et al., Phase noise influence in optical OFDM systems employing RF pilot tone for phase noise cancellation, Journal of Optical Communications, Vol.32(2), 141-145, 2011. DOI: 10.1515/joc.2011.017
[14] R. Maher, et al., Digital pulse shaping to mitigate linear crosstalk in Nyquist spaced 16QAM WDM transmission systems, OptoElectronics and Communication Conference, MO2B2, 2014. ISBN: 978-1-922107-21-3
[15] T. Xu, et al., Phase noise mitigation in coherent transmission system using a pilot carrier, Asia Communications and Photonics Conference, Proc. SPIE, Vol. 8309, 8309Z, 2011. DOI: 10.1117/12.904038
[16] G. Liga, et al., Digital back-propagation for high spectral-efficiency Terabit/s superchannels, Optical Fiber Communication Conference, W2A.23, 2014. DOI: 10.1364/OFC.2014.W2A.23
[17] S. J. Savory, Digital equalization in coherent optical transmission systems, Enabling Technologies for High Spectral-Efficiency Coherent Optical Communication Networks, John Wiley & Sons, Inc., 2016. DOI: 10.1002/9781119078289.ch8
[18] T. Xu, et al., Digital adaptive carrier phase estimation in multi-level phase shift keying coherent optical communication systems, International Conference on Information Science and Control Engineering, 1293-1297, 2016. DOI 10.1109/ICISCE.2016.276
[19] K. Kikuchi, Coherent transmission systems, European Conference on Optical Communication, Th.2.A.1, 2008. DOI: 10.1109/ECOC.2008.4729551
[20] N. A. Shevchenko, et al., Achievable information rates estimation for 100-nm Raman-amplified optical transmission system, European Conference on Optical Communication, 878-880, 2016. ISBN: 978-3-8007-4274-5
[21] R. Maher, et al., Linear and nonlinear impairment mitigation in a Nyquist spaced DP-16QAM WDM transmission system with full-field DBP, European Conference on Optical Communication, P.5.10, 2014. DOI: 10.1109/ECOC.2014.6963971
[22] T. Xu, et al., Analytical estimation in differential optical transmission systems influenced by equalization enhanced phase noise, Progress in Electromagnetics Research Symposium, 4844-4848, 2016. DOI: 10.1109/PIERS.2016.7735770
[23] D. K. Mynbaev, L. L. Scheiner, Fiber-optic communication technology, Prentice Hall, 2000. ISBN: 978-0139620690
[24] T. Okoshi, K. Kikuchi, Coherent optical fiber communications, Kluwer Academic Publishers, 1988. ISBN: 978-90-277-2677-3
[25] K. Kikuchi, Coherent optical communications - history, state-of-the-art technologies, and challenges for the future -, Opto-Electronics and Communications Conference, 1-

4, 2008. DOI: 10.1109/OECCACOFT.2008.4610574
[26] K. Kikuchi, History of coherent optical communications and challenges for the future, IEEE Summer Topical Meetings, TuC1.1, 2008. DOI: 10.1109/LEOSST.2008.4590512
[27] T. Xu, et al., Modulation format dependence of digital nonlinearity compensation performance in optical fibre communication systems, Optics Express, Vol.25(4), 3311-3326, 2017. DOI: 10.1364/OE.25.003311
[28] T. Xu, et al., Analytical BER performance in differential n-PSK coherent transmission system influenced by equalization enhanced phase noise, Optics Communications, Vol.334, 222-227, 2015. DOI: 10.1016/j.optcom.2014.07.094
[29] P. Bayvel, et al., Maximising the optical network capacity, Philosophical Transactions of the Royal Society A, Vol.374(2062), 20140440, 2016. DOI: 10.1098/rsta.2014.0440
[30] G. Jacobsen, et al., Error-rate floors in differential n-level phase-shift-keying coherent receivers employing electronic dispersion equalization, Journal of Optical Communications, Vol.32(3), 191-193, 2011. DOI: 10.1515/JOC.2011.031
[31] J. Kahn, K. P. Ho, Spectral efficiency limits and modulation/detection techniques for DWDM systems, Journal of Selected Topics in Quantum Electronics, Vol.10(2), 259-272, 2004. DOI: 10.1109/JSTQE.2004.826575
[32] R. A. Griffin, et al., 10 Gb/s optical differential quadrature phase shift key (DQPSK) transmission using GaAs/AlGaAs integration, Optical Fiber Communication Conference, WX6, 2002. DOI: 10.1109/OFC.2002.1036787
[33] S. Tsukamoto, et al., Coherent demodulation of 40-Gbit/s polarization-multiplexed QPSK signals with 16-GHz spacing after 200-km transmission, Optical Fiber Communication Conference, PDP29, 2005. DOI: 10.1109/OFC.2005.193207
[34] T. Xu, et al., Chromatic dispersion compensation in coherent transmission system using digital filters, Optics Express, Vol.18(15), 16243-16257, 2010. DOI: 10.1364/OE.18.016243
[35] G. Jacobsen, et al., Phase noise influence in coherent optical OFDM systems with RF pilot tone: IFFT multiplexing and FFT demodulation, Journal of Optical Communications, Vol.33(3), 217-226, 2012. DOI: 10.1515/joc-2012-0038
[36] E. Ip, J. M. Kahn, Digital equalization of chromatic dispersion and polarization mode dispersion, Journal of Lightwave Technology, Vol. 25(8), 2033-2043, 2007. DOI: 10.1109/JLT.2007.900889
[37] R. Maher, et al., Spectrally shaped DP-16QAM super-channel transmission with multi-channel digital back propagation, Scientific Reports, Vol.5, 08214, 2015. DOI: 10.1038/srep08214
[38] T. Xu, et al., Normalized LMS digital filter for chromatic dispersion equalization in 112-Gbit/s PDM-QPSK coherent optical transmission system, Optics Communications, Vol.283, 963-967, 2010. DOI: 10.1016/j.optcom.2009.11.011
[39] G. Jacobsen, et al., Phase noise influence in long-range coherent optical OFDM systems with delay detection: IFFT multiplexing and FFT demodulation, Journal of Optical Communications, Vol.33(4), 289-295, 2012. DOI: 10.1515/joc-2012-

0047

[40] G. Liga, C. B. Czegledi, T. Xu, PMD and wideband nonlinearity compensation: next bottleneck or fundamental limitation? European Conference on Optical Communication (ECOC), Workshop WS06, 2016. (Invited Talk)

[41] G. Jacobsen, et al., Receiver implemented RF pilot tone phase noise mitigation in coherent optical nPSK and nQAM systems, Optics Express, Vol.19 (15), 14487-14494, 2011. DOI: 10.1364/OE.19.014487

[42] T. Xu, et al., Analysis of chromatic dispersion compensation and carrier phase recovery in long-haul optical transmission system influenced by equalization enhanced phase noise, Optik, to appear, 2017.

[43] G. Jacobsen, et al., Capacity constraints for phase noise influenced coherent optical DnPSK systems, Progress in Electromagnetics Research Symposium, 140319195903, 2014. (Invited Talk)

[44] T. Xu, et al., Equalization enhanced phase noise in Nyquist-spaced superchannel transmission systems using multi-channel digital back-propagation, Scientific Reports, Vol. 5, 13990, 2015. DOI: 10.1038/srep13990

[45] T. Yoshida, T. Sugihara, K. Uto, DSP-based optical modulation technique for long-haul transmission, Next-Generation Optical Communication: Components, Sub-Systems, and Systems IV Proc. SPIE, Vol. 9389, 93890K, 2015. DOI: 10.1117/12.2078042

[46] T. Xu, et al., Quasi real-time 230-Gbit/s coherent transmission field trial over 820 km SSMF using 57.5-Gbaud dual-polarization QPSK, Asia Communications and Photonics Conference, AF1F.3, 2013. DOI: 10.1364/ACPC.2013.AF1F.3

[47] E. Ip, J.M. Kahn, Feedforward carrier recovery for coherent optical communications, Journal of Lightwave Technology, Vol. 25(9), 2675- 2692, 2007. DOI: 10.1109/JLT.2007.902118

[48] H. F. Haunstein, et al., Principles for electronic equalization of polarization-mode dispersion, Journal of Lightwave Technology, Vol. 22(4), 1169-1182, 2004. DOI: 10.1109/JLT.2004.825333

[49] T. Xu, et al., Mitigation of EEPN in long-haul n-PSK coherent transmission system using modified optical pilot carrier, Asia Communications and Photonics Conference, AF3E.1, 2013. DOI: 10.1364/ACPC.2013.AF3E.1

[50] G. Jacobsen, et al., Study of EEPN mitigation using modified RF pilot and Viterbi-Viterbi based phase noise compensation, Optics Express, Vol.21(10), 12351-12362, 2013. DOI: 10.1364/OE.21.012351

[51] M.G. Taylor, Phase estimation methods for optical coherent detection using digital signal processing, Journal of Lightwave Technology, Vol. 17(7), 901-914, 2009. DOI: 10.1109/JLT.2008.927778

[52] T. Xu, et al., Analytical investigations on carrier phase recovery in dispersion-unmanaged n-PSK coherent optical communication systems, Photonics, Vol.3 (4), 51, 2016. DOI: 10.3390/photonics3040051

[53] T. Xu, et al., Comparative study on carrier phase estimation methods in dispersion-unmanaged optical transmission systems, Advanced Information Technology,

Electronic and Automation Control Conference, to appear, 2017.
[54] I. Fatadin, D. Ives, S. J. Savory, Differential carrier phase recovery for QPSK optical coherent systems with integrated tunable lasers, Optics Express, Vol. 21(8), 10166-10171, 2013. DOI: 10.1364/OE.21.010166
[55] T. Xu, et al, Carrier phase estimation methods in coherent transmission systems influenced by equalization enhanced phase noise, Optics Communications, Vol.293, 54-60, 2013. DOI: 10.1016/j.optcom.2012.11.090
[56] G. Jacobsen, et al., EEPN and CD study for coherent optical nPSK and nQAM systems with RF pilot based phase noise compensation, Optics Express, Vol. 20(8), 8862-8870, 2012. DOI: 10.1364/OE.20.008862
[57] A. Ellis, et al., The impact of phase conjugation on the nonlinear-Shannon limit: the difference between optical and electrical phase conjugation, IEEE Summer Topicals Conference, 209-210, 2015. DOI: 10.1109/PHOSST.2015.7248271
[58] G. Jacobsen, Influence of pre- and post-compensation of chromatic dispersion on equalization enhanced phase noise in coherent multilevel systems, Journal of Optical Communications, Vol. 32(4), 257-261, 2011. DOI: 10.1515/JOC.2011.053
[59] G. Liga, et al., On the performance of multichannel digital backpropagation in high-capacity long-haul optical transmission, Optics Express, Vol. 22(24), 30053-30062, 2014. DOI: 10.1364/OE.22.030053
[60] T. Xu, et al., Receiver-based strategies for mitigating nonlinear distortion in high-speed optical communication systems, Tyrrhenian International Workshop on Digital Communications, IEEE Photonics in Switching, P2.2, 2015. (Invited Talk)
[61] R. I. Killey, et al., Experimental characterisation of digital Nyquist pulse-shaped dual-polarisation 16QAM WDM transmission and comparison with the Gaussian noise model of nonlinear propagation, International Conference on Transparent Optical Networks, Tu.D1.3, 2014. DOI: 10.1109/ICTON.2014.6876439
[62] S. Le, et al., Optical and digital phase conjugation techniques for fiber nonlinearity compensation, OptoElectronics and Communication Conference, 1-3, 2015. DOI: 10.1109/OECC.2015.7340113
[63] S. Haykin, Adaptive filter theory (5th Edition), Prentice Hall, 2013. ISBN: 978-0132671453
[64] T. Xu, et al., Digital compensation of chromatic dispersion in 112-Gbit/s PDM-QPSK system, Asia Communications and Photonics Conference, Proc. Vol. 7632, 763202, 2009. DOI: 10.1364/ACP.2009.TuE2
[65] T. Xu, et al., Variable-step-size LMS adaptive filter for digital chromatic dispersion compensation in PDM-QPSK coherent transmission system, International Conference on Optical Instruments and Technology, Vol. 7506, 75062I, 2009. DOI: 10.1117/12.837834
[66] www.diva-portal.org
[67] T. Xu, et al., Influence of digital dispersion equalization on phase noise enhancement in coherent optical system, Asia Communications and Photonics Conference, AS1C.3, 2012. DOI: 10.1364/ACPC.2012.AS1C.3

[68] S. J. Savory, Digital filters for coherent optical receivers, Optics Express, Vol. 16 (2), 804-817, 2008. DOI: 10.1364/OE.16.000804
[69] T. Xu, et al., Digital chromatic dispersion compensation in coherent transmission system using a time-domain filter, Asia Communications and Photonics Conference, 132-133, 2010. DOI: 10.1109/ACP.2010.5682798
[70] R. Kudo, et al., Coherent optical single carrier transmission using overlap frequency domain equalization for long-haul optical systems, Journal of Lightwave Technology, Vol. 27(16), 3721-3728, 2009. DOI: 10.1109/JLT.2009.2024091
[71] N. Benvenuto, G. Cherubini, Algorithms for communications systems and their applications, John Wiley & Sons, Inc., New York, 2004. ISBN: 978-0-470-84389-5
[72] D. Lowe, X. Huang, Adaptive overlap-add equalization for MB-OFDM ultra-wideband, International Symposium on Communications and Information Technologies, 644-648, 2006. DOI: 10.1109/ISCIT.2006.339826
[73] T. Xu, et al., Frequency-domain chromatic dispersion equalization using overlap-add methods in coherent optical system, Journal of Optical Communications, Vol. 32(2), 131-135, 2011. DOI: 10.1515/joc.2011.022
[74] O. W. Kwon, C. K. Un, J. C. Lee, Performance of constant modulus adaptive digital filters for interference cancellation, Signal Processing, Vol. 26(2), 185-196, 1992. DOI: 10.1016/0165-1684(92)90129-K
[75] J. Benesty, P. Duhamel, Fast constant modulus adaptive algorithm, IEE Radar and Signal Processing, Vol.138(4), 379-387,1991. DOI: 10.1049/ip-f-2.1991.0049
[76] T. Xu, et al., Analytical estimation of phase noise influence in coherent transmission system with digital dispersion equalization, Optics Express, Vol. 19(8), 7756-7768, 2011. DOI: 10.1364/OE.19.007756
[77] Y. Mori, et al., Unrepeated 200-km transmission of 40-Gbit/s 16-QAM signals using digital coherent receiver, Optics Express, Vol. 17(3), 1435-1441, 2009. DOI: 10.1364/OE.17.001435
[78] T. Xu, et al., Close-form expression of one-tap normalized LMS carrier phase recovery in optical communication systems, International Conference on Wireless and Optical Communications, Vol. 9902, 990203, 2016. DOI: 10.1117/12.2261932
[79] G. Goldfarb, G. Li, BER estimation of QPSK homodyne detection with carrier phase estimation using digital signal, Optics Express, Vol. 14(18), 8043-8053, 2006. DOI: 10.1364/OE.14.008043
[80] D. S. Ly-Gagnon, et al., Coherent detection of optical quadrature phase-shift keying signals with carrier phase estimation, Journal of Lightwave Technology, Vol. 24(1), 12-21, 2006. DOI: 10.1109/JLT.2005.860477
[81] www.mdpi.com
[82] A. J. Viterbi, A. M. Viterbi, Nonlinear estimation of PSK-modulated carrier phase with application to burst digital transmission, IEEE Transactions on Information Theory, Vol. 29(4), 543-551, 1983. DOI: 10.1109/TIT.1983.1056713
[83] T. Xu, et al., Analysis of carrier phase extraction methods in 112-Gbit/s NRZ-PDM-QPSK coherent transmission system, Asia Communications and Photonics Conference,